\RequirePackage{ifpdf}
\ifpdf % We are running pdfTeX in pdf mode
\documentclass[pdftex]{sigma}
\else
\documentclass{sigma}
\fi

\begin{document}

\renewcommand{\PaperNumber}{005}

\FirstPageHeading

\ShortArticleName{Andrew Lenard: A Mystery Unraveled }

\ArticleName{Andrew Lenard: A Mystery Unraveled}

% Names of the authors for the title of the paper
\Author{Jeffery PRAUGHT and Roman G. SMIRNOV~$^*$}
\AuthorNameForHeading{J. Praught and R.G. Smirnov}

\Address{Department of Mathematics and Statistics, Dalhousie University,\\
Halifax, Nova Scotia, Canada, B3H 3J5} 
\Email{\href{mailto:praughtj@mathstat.dal.ca}{praughtj@mathstat.dal.ca}, 
\href{mailto:smirnov@mathstat.dal.ca}{smirnov@mathstat.dal.ca}} 
\URLaddressMarked{\url{http://www.mathstat.dal.ca/~smirnov/}}

\ArticleDates{Received September 29, 2005, in final form October 03, 2005; Published online October 08, 2005}

 \Abstract{The theory of bi-Hamiltonian systems has its roots in what is commonly referred to as the ``Lenard
recursion formula''. The story about the discovery of the formula told by Andrew Lenard is the subject of this
article.}


\Classification{01A60; 35Q53; 35Q51; 70H06} 

\section{Introduction}

This aim of this review article is to present the untold story about the use of the name Lenard in
many concepts that form the backbone of bi-Hamiltonian (multi-Hamiltonian) theory. Originally, the
theory came to prominence with the fundamental 1978 paper by Franco Magri \cite{M78}, followed
almost immediately by the 1979 paper due to Israel Gel'fand and Irina Dorfman \cite{GD79} that
developed and extended the results presented in \cite{L76} and \cite{M78}. Since then, many
scientists have been working on the development of the theory of bi-Hamiltonian systems, making it
one of the most active areas of research in the field of   mathematical physics (see, for example,
\cite{AD05,B98,B96,BA92,D00,D91,FR94,FF81,GD79,GZ00,M78,M03,MM84,MMT88,O93,S97,S04}, 
as well as the relevant references therein).

The majority of the hundreds of papers written to date on the subject are invariably based on the
use of such concepts as the ``Lenard bicomplex'', ``Lenard chain'', ``Lenard recursion operator'',
``Lenard scheme'', and so forth (see, for instance, \cite{AD05, GZ00, M03,MMT88}). This leads one
to believe that Andrew Lenard must have made a fundamental contribution to the theory, yet, as
everybody working in the area knows, {\it no paper on the subject under his name has ever been
written}. Although  the papers by Clifford Gardner {\em et al} \cite{GGKM74} and Peter Lax
\cite{L76} contain short paragraphs that strongly allude to  Andrew Lenard's contribution, the
whole story told by him (see below) appears to be as fascinating as the result itself. Furthermore,
it must be said that the discovery of Lenard had fundamental consequences beyond the theory of
bi-Hamiltonian systems that it originated. The notion of a {\em recursion operator}, introduced by
Peter Olver in~\cite{O77}, is not, in general, a byproduct of the existence of two or more
Hamiltonian structures. Although the Lenard recursion operator for the Korteweg--de Vries equation
comes from the recursion relation~(\ref{L1}), more generally, a recursion operator is a property of
a symmetry group nature (see \cite{O77, O93} for more details), rather than the existence of a bi-Hamiltonian
structure.

In what follows, we reproduce\footnote{With A. Lenard's permission.} the story obtained by one of us
(JP) in full, preceded by a brief review of the mathematical background involved. We believe that
the results of this historical investigation will be of interest to the scientists working in the
area as well as anyone interested in the history of 20th century mathematics.

\section{The emergence of the theory}

 In the last forty years or so, the Korteweg--de Vries (KdV) equation has received much attention in the
mathematical physics literature following the pioneering work of Kruskal and Zabusky~\cite{ZK65} in
the mid-sixties.  As is well-known, in this work the authors have reported numerical observations
demonstrating that the KdV solitary waves pass through each other with no change in shape or speed.
The results presented in a series of papers by Gardner, Green, Kruskal, Miura and those that
followed \cite{GGKM67, M68, MGK, SG69, G71, FZ71, KMGZ70, GGKM74}, gave rise to the new {\em theory
of solitons} and indicated applications to many areas of mathematical physics that are still being
actively advanced today. Thus, for example, soon after the breakthrough of 1965, Gardner, Green,
Kruskal, and Miura \cite{GGKM67} enriched the theory with another fundamental development; it was 
a~new method later called  the {\em inverse scattering method} (ISM).  In a nutshell, the method
allows one to find the solution to the nonlinear problem of solving the KdV equation via a series
of linear computations.  Moreover, by using the ISM, as well as the new technique later called the
{\em Miura transform}, the authors demonstrated the existence of an infinite number of conservation
laws for solutions of the KdV equation and explicitly derived several of them \cite{MGK, M68}. In
turn, the existence of an infinite number of conserved quantities for the KdV equation was another
important step in advancing  the new theory.

Importantly, this discovery provided a framework for the introduction of a new Hamiltonian
formalism, suitable not only for studying the KdV equation but also other nonlinear partial
differential equations that exhibited similar properties (see \cite{D91} and the references
therein). It was first shown that the Hamiltonian formalism of classical mechanics could be
naturally incorporated into the study of the KdV equation. The consequences of this discovery,
reported in 1971, independently and almost simultaneously by Gardner \cite{G71} and Faddeev \&
Zakharov \cite{FZ71}, eventually reached far beyond the study of the KdV equation. In what follows
we reproduce the main features of the Hamiltonian formalism for the KdV equation and then show how
the Lenard recursion formula fits naturally within the general theory.

Consider the KdV equation of the following form:
\begin{gather}
     u_t = 6uu_x + u_{xxx}.
\label{1}
\end{gather}
 Gardner \cite{G71} and Faddeev \& Zakharov \cite{FZ71} observed that the right hand side of
 equation (\ref{1}) can be rewritten as follows:
\begin{gather}
\label{2}
          u_t = P_0 \frac{\delta H_0}{\delta u(x)},
\end{gather}
where $H_0$ is given by
\begin{gather}
     H_0 = \int \left(u^3 - \frac{1}{2}u_x^2 \right)dx,
\end{gather}
and
\begin{gather}P_0=\frac{\partial }{ \partial x}, \label{P}
\end{gather}
while the expression $\frac{\delta}{\delta u(x)}$ denotes the {\em gradient\/} (or, the {\em
Fr\`{e}chet derivative}). The gradient $\delta_u$ acts on any functional $F = F(u, u_x, u_{xx},
\ldots )$ as follows:
\begin{gather}
\label{Fr}
 \frac{\delta F}{\delta u} = \frac{\partial f }{\partial u} - \frac{\partial }{\partial x}\left(\frac{\partial
f}{\partial u_x}\right) + \frac{\partial^2}{\partial x^2}\left(\frac{\partial f}{\partial u_{xx}}\right) - \cdots,
\end{gather}
where $F (u, u_x, u_{xx}, \ldots )  = \int f (u, u_x, u_{xx}, \ldots )\,   \mbox{d}x.$

One immediately observes that formula (\ref{2}) bears a striking resemblance to the corresponding
formula for a Hamiltonian vector field $X_{H_0}$ in classical mechanics, namely
\begin{gather}
\label{X}
X_H = P_0\,\mbox{d}H_0
\end{gather}
 or, alternatively, $X_H = [P_0, H_0]$, where $[\cdot,\cdot]$ denotes the Schouten bracket. The quantities $P_0$ and
$H_0$ are  the corresponding Poisson bi-vector and Hamiltonian function, respectively, defined on a
finite-dimensional manifold $M$. The triple $(M, P_0, X_{H_0})$ is said to be a {\em Hamiltonian
system}. The most important feature of the Poission bi-vector that appears in formula (\ref{X}) is
that it can be used to define the corresponding Poisson bracket, which is the mapping $\{\cdot ,\cdot\}_0:$ 
${\cal F}(M)\times {\cal F}(M)$ $\rightarrow$ ${\cal F}(M)$ given by
\begin{gather}
\{f_1,f_2\}_0 := P_0\mbox{d}f\mbox{d}g,
\label{p1}
\end{gather}
 where $f_1,f_2 \in {\cal F}(M)$ and ${\cal F}(M)$ denotes the space of smooth functions on $M$. 
 The $\mathbb{R}$-bilinear mapping (\ref{p1}) is skew-symmetric and satisfies
the Jacobi identity. As is well-known, these key properties are central to the Hamiltonian
formalism of classical mechanics developed for finite-dimensional systems.  The most remarkable
observation that has been made independently in~\cite{G71} and \cite{FZ71} is that the differential
operator $P_0$ appearing in formula (\ref{2}) can be used to define a Poisson bracket as well.
Thus, in complete analogy with~(\ref{p1}) one defines:
\begin{gather}
 \{F_1,F_2\}_0 := \int \frac{\delta F_1}{\delta u(x)}\frac{\partial}{\partial x}\frac{\delta F_2}{\delta u(x)}\,
\mbox{d}x. \label{p2}
\end{gather}
 Remarkably, the bracket defined by (\ref{p2}) is also skew-symmetric and satisfies the Jacobi identity (see
\cite{L76} for the proof). Equation (\ref{1}) is rich in conserved quantities, three of which are
classical:
\begin{gather}
F_0  =   \int \frac{u}{2}\,\mbox{d} x, \nonumber\\
F_1  =   \int \frac{u^2}{2}\,\mbox{d} x, \nonumber\\
H_0 = F_2  =   \int\left(u^3 - \frac{u_x^2}{2}\right)\,\mbox{d} x.\label{cm}
\end{gather}
In view of formula (\ref{Fr}), their corresponding gradients are given by
\begin{gather}
 G_0 = \frac{\delta  F_0}{\delta u(x)}  =  \frac{1}{2}, \nonumber\\
G_1 = \frac{\delta  F_1}{\delta u(x)}  =   u, \nonumber\\
G_2 = \frac{\delta  H_0}{\delta u(x)} = \frac{\delta F_2}{\delta u(x)}  =  3u^2 + u_{xx}.
\label{Gs}
\end{gather}
 The following few ``simple'' formulas had a paramount impact on the development of a theory which had
ramifications and echoes in many areas of mathematics, including differential geo\-metry, the theory
of Lie groups, and Hamiltonian mechanics to name a few. The first observation is that the
representation of type (\ref{2}) for the KdV equation (\ref{1}) is not unique. For example, the
operator
\begin{gather}
\label{P1}
     P_1 = \frac{\partial^3}{\partial x^3} + 4u\frac{\partial}{\partial x}+2u_x
\end{gather}
can be used to define {\em another} Poisson bracket in much the same way as in (\ref{p2}):
\begin{gather}
\{F_1,F_2\}_1 := \int \frac{\delta F_1}{\delta u(x)}P_1\frac{\delta F_2}{\delta u(x)}\, \mbox{d}x.
\label{p3}
\end{gather}
\begin{figure}[ht]
    \centering
    \includegraphics[width=1.0\textwidth]{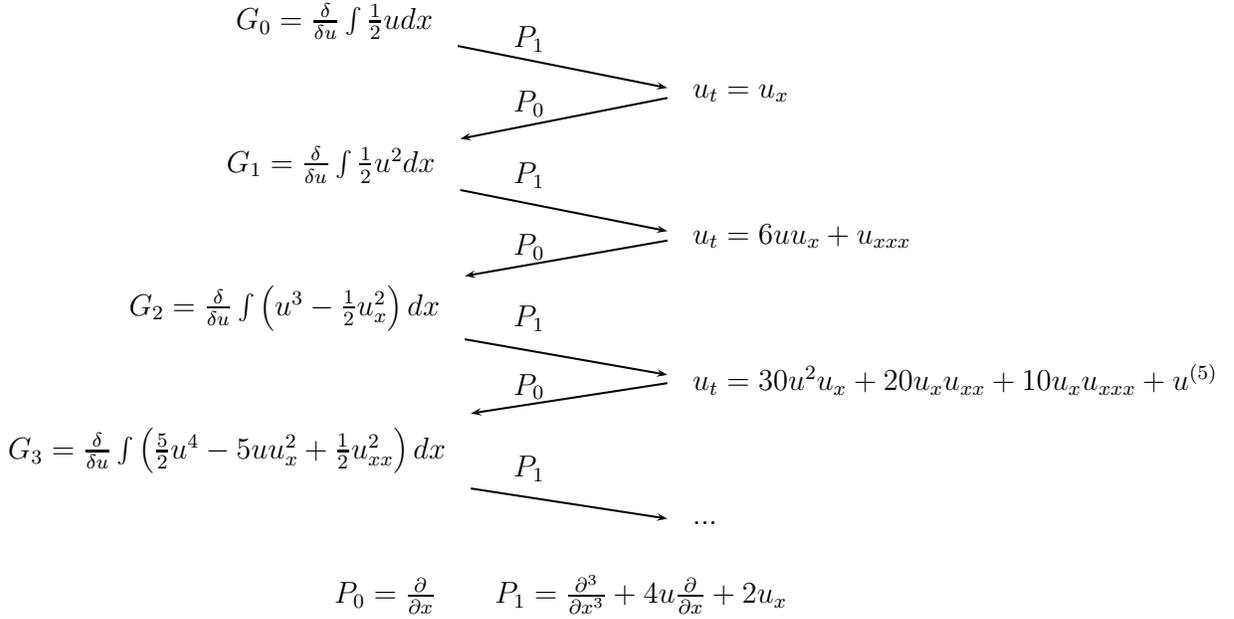}
      \caption{The Lenard recursion formula.}
      \vspace{-4mm}
\end{figure}

\noindent
Remarkably, the Poisson bracket $\{\cdot , \cdot\}_1$ is also skew-symmetric and satisfies the Jacobi
identity (see \cite{O93} for more details and proofs). Moreover, it can be matched with the
corresponding Hamiltonian $H_1$, that together with (\ref{P1}) yields a formula analogous to
(\ref{2}):
\begin{gather}
\label{kdv2}
u_t = P_1\frac{\delta H_1}{\delta u(x)},
\end{gather}
where $H_1$ is given by
\begin{gather}
\label{H1} H_1 = F_1 = \int \frac{u^2}{2}\,\mbox{d}x
\end{gather}
 and the right hand side of (\ref{kdv2}) is the same as the right hand side of (\ref{1}). Combining the 
 formulas~(\ref{2}) and (\ref{kdv2}), one easily arrives at the following important formula which can be
viewed as a precursor to the {\em Lenard recursion formula}:
\begin{gather}
P_1G_1 = \frac{\partial}{\partial x}G_2 = P_0G_2.
\label{L1}
\end{gather}
 Note that in view of (\ref{p2}), (\ref{p3}), and (\ref{L1}), the functionals $F_1$ and $F_2$ (having the
gradients $G_1$ and~$G_2$ respectively) are in involution with respect to both $\{\cdot, \cdot\}_0$ 
and $\{\cdot, \cdot\}_1$:
\begin{gather}
\label{p12}
\{F_1, F_2\}_0 = \{F_1, F_2\}_1 = 0.
\end{gather}
 Furthermore, Lax proved in \cite{L76} the existence of more of these conserved functionals $F_n$, $n=0, 1, 2,
\ldots $  exhibiting the property (\ref{p12}). His proof is based on a generalization of the relation~(\ref{L1})
which is nothing but the celebrated {\em Lenard recursion formula}:
\begin{gather}
P_1G_n = \frac{\partial}{\partial x}G_{n+1} = P_0G_{n+1},
\label{L2}
\end{gather}
 where the $G_n$'s  are the gradients of the conserved functionals $F_n$'s. It is easy to check that $G_0,$
$G_1$ and $G_2$ given by (\ref{Gs}) satisfy the relation (\ref{L2}). Furthermore, it easily follows from
(\ref{L2}) that the functionals $F_0, F_1, F_2, \ldots $ corresponding to the gradients  $G_0, G_1, G_2, \ldots
$ generated via (\ref{L2}) are mutually in involution with respect to both Poisson brackets $\{\cdot,\cdot\}_0$ and
$\{\cdot,\cdot\}_1$. In addition to the above, the Lenard recursion formula (\ref{L2}) has a number of important
consequences, among which  we single out the following two:
\begin{itemize}
\itemsep=0pt

\item The existence of two Hamiltonian representations for the KdV equation (\ref{1}) gives rise to an infinite
sequence of conserved functionals of (\ref{1}).

\item For every $n \ge 1$, the Lenard recursion formula (\ref{L2}) defines a higher order KdV equation,
which has the same conserved quantities as the basic KdV equation (\ref{1}). Therefore, the Lenard
recursion formula leads to the {\em KdV hierarchy.}

\end{itemize}

% new page command necessary to put text before diagram on this page, otherwise the diagram would show first
% and the text "We proceed ..... Lenard recursion formula" would appear below it.
%\newpage

We proceed with the following diagram (see Figure 1), which illustrates the properties and consequences of the
Lenard recursion formula.

 Shortly after this discovery concerning the KdV equation, it was shown by Magri \cite{M78} that the property of
having two Hamiltonian representations was not a specific feature of the KdV equation alone, but
rather a general property that could be found for other nonlinear PDEs with the same remarkable
consequences. In his celebrated 1978 paper \cite{M78}, Magri studied from this viewpoint, the Harry
Dym and the modified KdV equations, thus developing a general scheme for studying these soliton
equations as bi-Hamiltonian systems. A year later, in another fundamental paper, Gel'fand and
Dorfman \cite{GD79} extended these ideas to the field of finite-dimensional Hamiltonian systems.
One cannot help noticing that Hamiltonian formalism originated in classical mechanics and was then
applied to the study of soliton equations, while the bi-Hamiltonian formalism travelled in the
opposite direction.

And the story began \dots

\section{Andrew Lenard's story}

 In this section we reproduce the story told by Andrew Lenard describing the events preceding and following the
discovery of the Lenard recursion formula. In order to make the exposition  clearer, we refer
throughout the story to the corresponding references or the formulas presented in  the previous
section. Here is the story.

\medskip

\centerline{ * * * }

 ``Thank you for your communication. It is quite appropriate in the connection of your work, and I shall try to
reply as best as I can.

 In the earlier part of the 1960s, I was a scientific staff member of the Plasma Physics Laboratory (PPL)
operated by Princeton University in conjunction with the Atomic Energy Commission. There, Martin
Kruskal was a friend and colleague. He, together with his co-worker Norman Zabusky, discovered an
astonishing phenomenon of the KdV differential equation, not until then noticed; namely, that in
spite of its non-linear nature, certain wave solutions maintained their shapes unchanged after
passing through a time interval of intense non-linear interaction\footnote{The author refers to the results
presented in \cite{ZK65}.}. This was followed by the discovery of a type of
``linearization'' of the problem by a functional transformation relating it to the 1-dimensional
Schr\"{o}dinger Equation\footnote{That is, the inverse scattering method \cite{GGKM67}.}. In
addition, first one and then several simply expressible constants of motion\footnote{See (\ref{cm}).}
were found for the KdV evolution equation. Due to the combined work of Clifford Gardner,
John Greene, and others, soon an infinite hierarchy of such constants of motion were generated\footnote{See
the references \cite{GGKM67, MGK, M68}.}.

 I left the PPL at this point to come to Indiana University. However, on a visit back to Princeton during the
summer of 1967 (I believe) I went back to the PPL to see my old friends. It was there that something remarkable
happened.

 I arrived at coffee time in the afternoon. In the common room there were some blackboards. In front of one a
crowd was gathered, centered around Kruskal, excitedly discussing something. I went up to ask what
it was. They explained that another differential equation, similar to KdV but of higher order, was
found, showing all those features of KdV I just described\footnote{See Figure 1.}. Someone wanted
to know how one could {\em systematically} discover it, rather than just by lucky hit and miss. I
heard Martin Kruskal shout at me: ``There must be a method to generate many more, probably
infinitely many such higher and higher order DE's, don't you think, Andrew?''

 I asked for a yellow pad and pen, and went to sit down in a quiet corner to gather my thoughts. I was at that
point particularly expert on generating functions as a means of summarizing information about
infinite sequences in one mathematical construct. So naturally I tried this idea on the problem at
hand, and it worked! It took me only {\em fifteen minutes or so}, and I could explain to the
gathered friends how by means of a generating function an infinite hierarchy of KdV-like DE's could
be generated, all of them having the same kind of behavior\footnote{See Figure 1.}.

This was greeted with admiration and satisfaction. I had my coffee and left.

 Later, I saw that an article in the Comm.\ Appl.\ Math.\ (Courant Institute, NYU) by Gardner, Greene and Miura and
perhaps others, had an appendix on my discovery\footnote{See the reference \cite{GGKM74}.}. I myself {\em
never published anything, nor concerned myself with the subject, then or since}.

 Several times during the intervening years I was surprised to hear my name being mentioned in connection with
this, but actually much of it in connection with mathematics too high for me to appreciate. For instance,
someone once told me that what I discovered was a dynamical system on a symplectic manifold with {\em two}
different Hamiltonian structures\footnote{See the formulas (\ref{2}) and (\ref{kdv2}) as well as, for example,  the
references \cite{GD79, MM84, O93} for more details.}. And someone mentioned the ``Lenard-Recursion
Operator''\footnote{See \cite{O93} for more details.},  and asked whether that was the same person as~I.

 Naturally, I am satisfied that I could make a contribution, even in such a fortuitous and judicious manner as I
told you.

 I hope this story will be satisfactory for you and answer your questions. By all means, feel free to share it
with any like minded person if you care to. I don't mind it at all if the history of how ``Lenard'' became a
concept in this area will be generally known.

Much good luck for your own studies, and sincerely yours:

\hspace*{10cm}Andrew Lenard

\bigskip

\noindent
 P.S. I recall that a mathematician at Dalhousie University (probably retired by now) whom I knew as a friend
and colleague when he was at Indiana University during the early 1970s, is Peter Fillmore. Say hello to him for
me if you see him.''

\medskip

\centerline{ * * * }

\subsection*{Acknowledgements}

We wish to thank Peter Olver for his careful reading of the manuscript and constructive comments.
The first author (JP) would like to express his gratitude to fellow graduate students Caroline Adlam,
Denis Falvey, Josh MacArthur, John Rumsey and Jin Yue for the stimulating atmosphere and useful
discussions. The research was supported in part by a National Sciences and Engineering Research
Council of Canada Discovery Grant (RGS).

\LastPageEnding

\end{document}